\documentclass[pra,onecolumn,preprintnumbers,amsmath,amssymb,superscriptaddress,longbibliography,nofootinbib]{revtex4-2}

\usepackage{graphicx}
\usepackage{dcolumn}
\usepackage{bm}
\usepackage{comment} 
\usepackage{float}
\usepackage{bigints}
\usepackage[utf8]{inputenc}
\usepackage{esint}
\usepackage{wrapfig}
\usepackage{lmodern}
\usepackage{amsfonts}
\usepackage{epstopdf}
\usepackage[arrowdel]{physics}
\usepackage[T1]{fontenc}
\usepackage{amsmath,amssymb,amsthm}
\usepackage{tikz}
\usepackage{cancel} 
\usepackage{color}
\usepackage{subcaption}
\usepackage[shortlabels]{enumitem}
\usepackage{listings}
\usepackage{xcolor} 
\usepackage{hyperref} 
\usepackage{breqn}
\usepackage{booktabs}
\usepackage{blindtext}

\newcommand{\cor}[1]{{#1}}

\begingroup\makeatletter
\catcode`,=\active
\global\let\breqn@comma,
\protected\gdef,{\ifmmode\expandafter\breqn@comma\else\expandafter\active@comma\fi}
\endgroup

\begin{document}

\title{Addressing the Non-perturbative Regime of the Quantum Anharmonic Oscillator by Physics-Informed Neural Networks}

\author{Lorenzo Brevi}
\author{Antonio Mandarino}%
\author{Enrico Prati}
 \email{enrico.prati@unimi.it}
\affiliation{Department of Physics Aldo Pontremoli,Università degli Studi di Milano, Via Celoria 16, 20133 Milano, Italy}

\date{\today}

\begin{abstract}
The use of deep learning in physical sciences has recently boosted the ability of researchers 
to tackle physical systems where little or no analytical insight is available. 
Recently, the Physics$-$Informed Neural Networks (PINNs) have been introduced as one of the most promising tools  
to solve systems of differential equations guided by some physically grounded constraints. 
In the quantum realm, such an approach paves the way to a novel approach to solve the Schr\"odinger equation for 
non-integrable systems. 
By following an unsupervised learning approach, we apply the PINNs to the anharmonic oscillator in which 
an interaction term proportional to the fourth power of the position coordinate is present. 
We compute the eigenenergies and the corresponding 
eigenfunctions while varying the weight of the quartic interaction.
We bridge our solutions to the regime where both the perturbative and the strong coupling theory work, including the pure quartic oscillator.
We investigate systems with real and imaginary frequency, laying the foundation for 
novel numerical methods to tackle problems emerging in quantum field theory. 
\end{abstract}

\maketitle

\section{Introduction}
Much of the modern understanding in physics is due to the theoretical framework established by by quantum field theories. 
Ranging from the quantization of the electromagnetic field to the explanation of heat conduction in solids, 
the ingenious and \textit{revolutionary trick} was to rethink the fields as a collection of harmonic oscillators \cite{Altland_Simons_2010}.
Since harmonic systems are one of the few integrable systems either in classical or quantum mechanics, 
and even if the solution of harmonic systems is an academic exercise, 
it is also wide knowledge how predictive these systems are.  
This implies that analytical computations allow the knowledge of their entire spectrum 
and to gain insight from a multitude of physical systems thanks to the mapping between 
bosonic systems and the quantum harmonic oscillator \cite{coleman2019quantum}. 
This boosted not only the acclaimed achievements in the quantum theory of radiation \cite{Glauber_nob,Haroche_nob},
but also the application of field theoretical tools to treat the small oscillations in molecules \cite{molecules}
or in the solid lattice and fostered our understanding of condensed matter. \cite{ashcroft2021solid}. 

However, if the harmonic potential is modified by the addition of terms proportional to higher powers of 
the position coordinates, the system almost always loses its integrability
\footnote{It is worth noting that the Morse potential (and several of its variations) 
can account for the anharmonic behavior and are still integrable.} \cite{Turbiner_book}.   
Nevertheless, since the pioneering work by Lord Rayleigh in classical physics \cite{rayleigh1896theory} 
and subsequently by Schr\"odinger \cite{Shrod_pert} in the quantum context, 
it was shown that slight perturbation breaking the harmonicity can be successfully treated within the framework of perturbation theory. 
In the present paper, we aim to provide a deep-learning perspective on a system that has long served  
as a prototypical interaction model \cite{BenderWu69,BenderWu74}, and it served and still serves 
to study innovative techniques in quantum field theory \cite{BenderWu68, PRD0,PRD1,PRD2,PRD3}.
The anharmonic oscillator with a term proportional to $x^4$ can be indeed seen as a $\lambda \phi^4$ theory in $(0 + 1)-$dimensions.
Furthermore, such a system has been consistently used in several contexts, most notably in 
solid-state physics to account for the hardening of the phonon dispersion relation \cite{PRB0,PRB1, PRLquartPhonon},  
to investigate the formation of molecules with hydrogen bonds and their vibrational modes \cite{Chem1, Chem2}, and also for foundational studies such as 
nonlinear quantum mechanics \cite{bialynicki1976nonlinear, dekker1981classical}, quantum chaos and control \cite{quartic_chaos, PRLappQuart}.

Some of us have already developed both supervised \cite{maronese2022quantum,moro2023anomaly,corli2023max} and unsupervised \cite{rocutto2021quantum} machine learning methods, including supervised quantum machine learning to address ground state classification \cite{lazzarin2022multi, LMG_grossi, Monaco_ANNNI} of a physical system. Here, we aim to address the unsupervised solution of the Schr\"odinger \cite{PINNschrod} equation, to explore the potential of Physics-Informed Neural Networks (PINNs) for solving Partial Differential Equations (PDEs) as a paramount representative application to nontrivial physical phenomena \cite{RAISSI2019686}. 

We applied the method to find both the ground state and the excited states 
of the quantum harmonic and anharmonic oscillator, with real or imaginary frequency.
We employed a PINN that splits into two separate networks \cite{harcombe2023physicsinformed}, one for the eigenvalue and one for the eigenfunction. Furthermore, it also employs an auxiliary output \cite{YUAN2022111260} to obtain the normalization integral. We also employed transfer learning, utilizing the weights and biases of the already trained networks to obtain an excellent initialization for networks with similar parameters \cor{of the system Hamiltonian}.

The remaining part of the paper is organized as follows. In Section \ref{sec:back}, we briefly introduce the details of the deep learning model employed, explaining in particular loss functions and architecture of the neural network. 
 Section \ref{sec:net} gives the detail of the quantum anharmonic oscillator, 
 and delves into the detailed explanation of the unsupervised learning strategy that we followed. Section \ref{sec: harm_osc}  shows the results for the unperturbed (harmonic) oscillator while in Section \ref{sec:anharm} the results for the anharmonic oscillator are described. Section \ref{sec:pureq} shows the results for the pure quartic oscillator while in Section \ref{sec:ener} we address the scaling behavior of the eigenenergies as a function of $\lambda$, \cor{the interaction strength}. Lastly, Section \ref{sec:concl} concludes the paper and address some interesting perspectives for future research directions. 

\section{Background} 
This sections contain in a concise form the required background information for our analysis. 
We start with a brief introduction to the considered neural network architecture, 
afterward we discuss the relevance of the considered model, especially from a field theory point of view. 
\label{sec:back}
\subsection{Physics-informed neural networks}
Physics-informed neural networks \cite{RAISSI2019686}, or PINNs are a category of neural networks that aim to solve Partial Differential Equations by encoding prior knowledge about the physical system, most importantly the differential equation itself, in their loss function. This loss can be written as:
\begin{equation} \label{eq: loss_generic}
    f = \cor{\mathcal{L}}_{PDE} + \cor{\mathcal{L}}_{data} + \cor{\mathcal{L}}_{phys}
\end{equation}
Where the term \(\cor{\mathcal{L}}_{data}\) is the standard loss for a Neural Network \footnote{In our case this loss will not be present since we are working in an unsupervised setting.}. \(\cor{\mathcal{L}}_{phys}\) corresponds to the losses given by the physical constraints of the system, like boundary conditions and initial conditions.
Lastly, \(\cor{\mathcal{L}}_{PDE}\) is the mean squared error of the differential equation. For instance, given a PDE with implicit solution \(\mu(t)\) 
that depends on \(\mu(t)\) and its derivative \(\partial_t\mu(t)\) with respect to the independent variable $t$, the differential equation can be written as:
\begin{equation} \label{eq: generic_PDE}
    \partial_t\mu(t) + N[\mu;\xi] = 0
\end{equation}
where $N$ is a nonlinear operator and \(\xi\) is some set of parameters. Given this, we can write \(\cor{\mathcal{L}}_{PDE}\) for a network designed to solve this equation as:
\begin{equation} \label{eq: generic_PDE_loss}
    \frac{1}{n}\sum_{i=1}^n(\partial_t\cor{\mu_{net}}(t_i) + N[\cor{\mu_{net}}(t_i);\xi])^2
\end{equation}
Where the \({t_i}\) is the set of $n$ points, called collocation points, where the PDE will be evaluated.
This makes it possible to train neural networks even in the presence of noisy or very little data, and even in a fully unsupervised setting, such as in this paper. These networks have had a number of applications in various fields. Just to name a few applications of PINNs, the most recent and relevant use cases are in fluid dynamics \cite{PhysRevFluids.4.034602, Raissi20201026, PhysRevFluids.6.073301}, heat transfer problems \cite{Cai2021}, scattering in composite optical materials \cite{Chen202011618}, oscillating dynamics of black holes \cite{PhysRevD.107.064025}, reconstruction of top quark kinematic properties \cite{PhysRevD.107.114029} and to compute the electronic density in complex molecular systems \cite{PhysRevLett.125.206401}.

\subsection{Related works}
\cor{The Schr\"odinger equation stands at the cornerstone of the modern quantum theory and its subsequent 
development that led to the first and second quantum revolutions. 
Alas, the cases in which it is analytically solvable are only a few, and 
the central potential problem stands among them. This boosted the necessity to look for alternative ways of solving the differential equation giving the dynamics of a quantum object. 
In recent years, the enthusiasm around the potential of artificial intelligence methods triggered their 
applications also in quantum science and technology (see \cite{Carleo_RMP, krenn2022scientific} for thorough discussions).
Several attempts have already been made to address the Schr\"odinger equation via Artificial Neural Networks. For instance, the authors in \cite{LAGARIS19971} already had some success in 
solving the dynamics of a system under the action of the Morse potential, anharmonic but still integrable. 
When moving to the particular class of neural networks considered here, unsupervised PINNs have already proven to be successful in addressing simple systems \cite{PINNschrod} as well as the Schr\"odinger equation with random potentials \cite{harcombe2023physicsinformed}. Moreover, PINNs have also been used to study the Schr\"odinger equation with time-dependent potentials \cite{timedep} and to obtain solutions of the nonlinear one that arises in wave-guided quantum optics \cite{9861710, 10.1063/5.0086038, li2022mix}. Deep complex neural networks \cite{deepcomplex} are also of special interest for future applications, due to their ability to naturally address complex-valued wavefunctions. They also have been used to discover the governing PDE of a system from data \cite{deepcomplex2}.}

\subsection{The quantum anharmonic oscillator}
The anharmonic oscillator model has played a pivotal role in many research areas in physics, mostly due to its 
simplicity but especially because it is a nontrivial nonlinear problem. It appeared to be crucial in the 
development of several branches of physics ranging from high-energy particle physics to low-energy molecular dynamics. 

\cor{Before delving into the discussion of our main problem, we review briefly the terminology. 
The solution of the Schr\"odinger equation provides the possible energy levels and the corresponding wave functions, describing the dynamics of a quantum object in a given potential field. 
The time-independent Schr\"odinger Equation allows to find the the stationary states of a quantum system, namely those for which the energy remains constant over time. To write it down, one needs to introduce the Hamiltonian operator \( \hat{H} \) that represents the total energy of a quantum system, such that: 

\begin{equation}
 \hat{H} \psi(\mathbf{r}) = E \psi(\mathbf{r})   
\end{equation}

\( \psi(\mathbf{r}) \) is the wave function that describes the spatial distribution of the particle, and \( E \) is the energy eigenvalue associated with that state.

The Hamiltonian operator of a particle of mass $m$ is usually decomposed into two main parts: the kinetic energy operator \( \hat{K} = \frac{\hat{\vec{p}}^{\,2}}{2 m} \) and a time-independent potential energy operator \( \hat{V} \). Specifically, the Hamiltonian can be written as:
\[
\hat{H} = \hat{K} + \hat{V} = -\frac{\hbar^2}{2m} \nabla^2 + V(\mathbf{r}), 
\]
where we have substituted the momentum operator $\hat{\vec{p}}$ with the corresponding a differential operator 
in the position representation, i.e., $\hat{\vec{p}} \rightarrow - i \hbar \nabla$ .
From now on for the sake of simplicity, we will omit the hat symbol on the operators.}

The Hamiltonian accounting for the total energy of the anharmonic oscillator in one-dimension 
for a system of unitary mass is: 
\begin{equation} \label{eq:Hamiltonian}
     H = \frac{p^2}{2} + V(x) = \frac{p^2}{2} + \frac{\omega^2 x^2}{2} + \lambda x^4
\end{equation}
in which $\omega$ is the harmonic frequency and $\lambda$ quantifies the strength of the interaction-like term, 
and finally $p$ and $x$ are the standard position and momentum operators in one dimension satisfying $[x, p] = i$. 
The potential $V(x)$ could assume two relevant shapes for physical applications, 
depending on the signs of $\omega^2$ and $\omega^2/\lambda$. 
When both terms are positive the potential has only one minimum and its shape is that of a ``super parabola'', 
whereas when the frequency is imaginary, we are in the presence of a double-well potential that is the archetype of a model exhibiting a spontaneously symmetry-breaking mechanism. 
Without losing generality in the following we will assume that the oscillator has unitary mass and we will work in natural units $m=c=\hbar=1$, 
where $m$ is the mass of the oscillator, $c$ and $\hbar$ are the speed of light in the vacuum and the reduced Planck constant respectively. 

Since the preliminary works of Bender and Wu \cite{BenderWu68, BenderWu69, BenderWu74} based on perturbative expansions in $\lambda,$ 
several techniques have been deployed to study the spectrum of this Hamiltonian. 
A notable observation noted by Bender and Wu, employing the Wentzel–Kramers–Brillouin approximation, 
was that for any $\lambda \neq 0$ the perturbative series is divergent.  
Later on Zinn-Justin and collaborators proposed perturbative 
solutions in terms of instantons connecting the two minima of potentials characterized by $\omega^2 < 0$, 
which are particularly suitable for weakly interacting scalar fields exhibiting spontaneously broken symmetries \cite{ZJ_1, ZJ_2, ZJ_3, ZJ_4}.
More recently, the Schr\"odinger equation stemming from the Hamiltonian in Eq. \eqref{eq:Hamiltonian} was mapped in a Riccati form 
and some approximated eigenfunctions were obtained in \cite{Turbiner_1, Turbiner_book}.

\section{Methodology} \label{sec:net}

To obtain eigenfunctions and eigenstates for the anharmonic oscillator, we employ two neural networks that will be trained in parallel. \cor{The first and main network takes as input the position $x$}, which will be sampled in an interval \([-\frac{L}{2}, \frac{L}{2}]\) where $L$ is a hyperparameter that controls the size of the computational domain and has to be chosen large enough that the wave function is vanishingly small at the system's boundary, but not so large that the network has to predict a vast area of just zeros.\\ \cor{It} will then output the wave function \(\psi(x)\) and the auxiliary output \(\nu(x)\), which is defined as:
\begin{equation}
\label{eq: integral}
    \nu(x) = \int_{-\frac{L}{2}}^x |\psi(k)|^2\, dk\
\end{equation}
\\
 and will be used to ensure normalization. On the other hand, \cor{the second network takes as input just the constant value $1$ and outputs a guess for the energy}. \\
We need to optimize a total of seven losses to train the network\cor{s}. First of all, we need a way to compute the integral of the squared modulus of the eigenfunction to be able to normalize the wave function. To do it in a mesh-free way we utilize an auxiliary output, \(\nu(x)\), and train the network to make it a guess for integral from zero to $x$ of the squared modulus of the outputted wave function, (see equation \eqref{eq: integral}). To ensure that this auxiliary output has the desired form, we utilize automatic differentiation to obtain the partial derivative of the auxiliary output, and require it to be equal to the squared modulus of the wave function, thus obtaining the integral loss:
\begin{equation} \label{eq: integ_loss}
    \left. \frac{\partial \nu(x')}{\partial x'}\right|_{x' = x} = |\psi(x)|^2,
\end{equation}
where the partial derivative of the auxiliary output can easily be obtained thanks to automatic differentiation. Of course, if this loss is satisfied then the auxiliary output will have the desired form \eqref{eq: integral}. At this point, we need to ensure that the output of the network respects the physical constraints of the system. This is accomplished through three more losses. The first is the normalization loss:
    \begin{equation} \label{eq: norm_loss}
    \begin{aligned}
        \nu(-\frac{L}{2}) = 0, \quad         \nu(\frac{L}{2}) = 1.
    \end{aligned}
    \end{equation}
which ensures that the wave function is normalized. Furthermore, and maybe more importantly, this loss forces the neural network to actually predict the eigenfunctions instead of just a constant stream of zeros, since it is the only loss to not be exactly satisfied when \(\psi(x) = 0\) for all $x$.\\
We then utilize the boundary condition loss:
    \begin{equation} \label{eq: bc_loss}
    \begin{aligned}
        \psi(-\frac{L}{2}) = 0, \quad    \psi(\frac{L}{2}) = 0
    \end{aligned}
    \end{equation}
to ensure that the wavefunction is vanishingly small at the borders of the domain. The last physical constraint to encode as a loss is the orthogonality of the eigenstates, which will allow the PINN to also find the excited states by imposing for the state predicted by the current network to be orthogonal to all previously found states. This can be accomplished by the orthogonality loss:
    \begin{equation} \label{eq: orth_loss}
        \sum_i\braket{\psi(x)}{\psi_i(x)}{} = 0
    \end{equation}
where \(\psi_i(x)\) are the known eigenfunctions. \cor{For the case of the ground state the orthogonality loss will always be zero. This is because there are no discovered states to compare the first output to.} Each time a network converges it will be saved and utilized to obtain the \(\psi_i(x)\) utilized to compute this loss in the training of the networks of higher energy states. At this point, we need to include the loss that will allow the network to gauge the correctness of the outputted eigenstate and eigenvalue for the current potential, that is the differential equation loss:
\begin{equation} \label{eq: eq_loss}
    -\frac{1}{2}\frac{\partial^2 \psi(x)}{\partial x^2} + (\frac{1}{2} \omega ^ 2 x ^ 2 + \lambda x ^ 4)\psi(x) - E \psi(x) = 0 
\end{equation}
which is just the Schr\"odinger equation that stems from the Hamiltonian in Eq. \eqref{eq:Hamiltonian}. 
At this point, we have a set of losses that will result in a neural network that will output one of the eigenstates for the given potential. However, we have no guarantee that this state is the one with the lowest energy among the ones that have not yet been discovered. To ensure this we need to add a condition that forces the network to minimize the energy. We use:
    \begin{equation}\label{eq:loss_en}
    \cor{e^{a(E_{PINN} - E_{init})}}
    \end{equation}
where \(E_{PINN}\) is the predicted energy, $a$ is a hyperparameter, and \(E_{init}\) is either another hyperparameter, when $n = 0$, or the energy of the previous state, for excited states. The choice to increase \(E_{init}\) as the quantum number increases has been taken to avoid an excessive weight for this loss for a higher energy state, which might cause the network to fail to converge at all. \cor{We employ the exponential of the difference of the $E_{PINN}$ and $E_{init}$ because it is a monotonically increasing positive function in $[-\infty, +\infty]$. This guarantees a loss that is both always positive and always pushes the network to reduce the energy. Thus, it should lead the network to discover the minimal energy state}. The energy minimization loss, however, will be greater than zero even when we reach the correct solution, and that will cause the network to converge to an incorrect state. In order to avoid this we gradually decrease this loss' weight as the training continues. In this way in the earlier epochs, this loss will steer the network towards lower energy states, while for the latter epochs it will become negligible and allow the network to settle to the correct solution.

The last loss that we implement is one that is not actually needed to obtain a solution, but is instead added to allow the network to converge much faster, acting as an inductive bias: the symmetry loss. Since the potential is symmetric with respect to 0, the wave functions that solve Eq. \eqref{eq: eq_loss} will also have to be alternatively symmetric and anti-symmetric with respect to 0. We can enforce this condition by adding the \cor{condition}:
    \begin{equation}
    \label{eq: symm-loss}
    \psi(x)-s\psi(-x) = 0
    \end{equation}
where $s$ is a parameter that assumes the value one if we are looking for a symmetric state ($n$ even), 
or $s=-1$ if we are looking for an anti-symmetric one ($n$ odd). For the sake of clarity, it is worth mentioning that 
the value of s starts at one and is switched every time a new model is trained.

\begin{figure}[ht] 
    \centering
    \includegraphics[width=0.9 \linewidth]{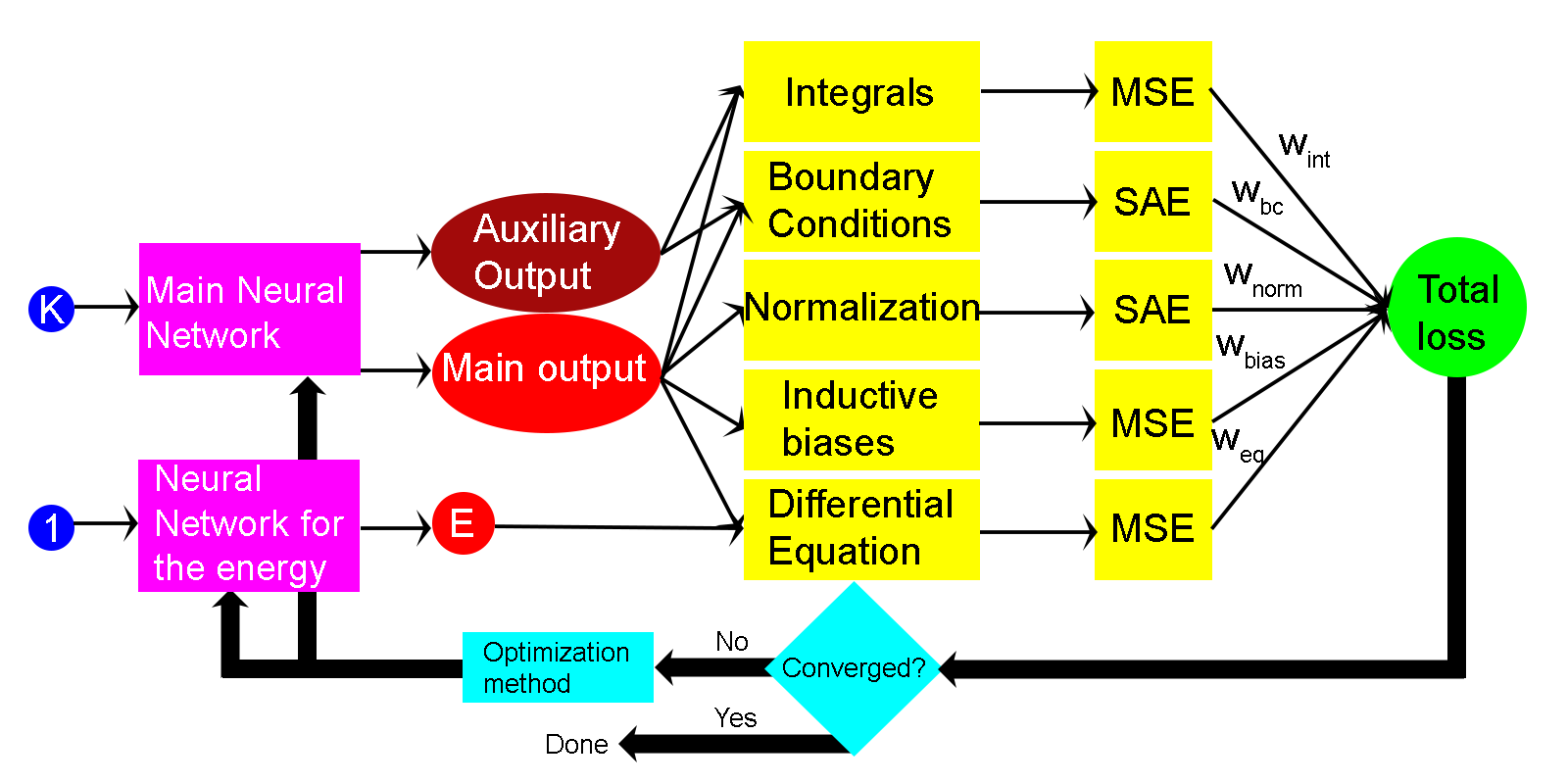}
    \caption{\cor{Graphical representation of the approach followed in this paper. The circles represent variables, the rectangles represent operations. The variables are also represented by primary colors, while the methods are represented by the secondary colors obtained by combining the colors of their input and their output. In blue there are the neural network inputs. In purple there are the two neural networks, the main one that computes the eigenstates and the one that computes the energy. In red there are the outputs of the neural networks: the auxiliary outputs, the main outputs which is the eigenfunction and the eigenvalue $E$. In yellow there are the operations needed to calculate the losses. The losses are the integral loss, the boundary condition loss, the normalization loss, the inductive biases and the differential equation loss. SAE is the Sum of Absolute Errors, SSE is the Sum of Squared Errors. In green there is the final loss. It is calculated by summing the partial losses weighted by empirically adjusted hyperparameters $w_{int}$, $w_{bc}$, $w_{norm}$, $w_{bias}$ and $w_{eq}$. In cyan there are the criteria and the optimization method.} }
    \label{fig:graph}
\end{figure}

For the architecture, the \cor{main} network\cor{, that is,} the one that outputs the wave function and the integral, is made up of seven hidden layers of $256$ neurons each. \cor{The one used to predict the energy, on the other hand, is made up of four hidden layers also of $256$ neurons each. Even if this network only has to perform a linear transformation, empirically the presence of hidden layers seems to vastly improve convergence}. Batches of $512$ collocation points were employed. We want each batch to span the whole domain, but we also want to fully utilize the mesh-free nature of the neural method. To do so we construct a grid of $512$ equally spaced points, and for each of them, we sample randomly in an interval centered on it, thus obtaining a batch that spans the whole domain but whose values change slightly at each iteration, ideal characteristics to optimize the network's convergence. \cor{For the activation function, we employed the hyperbolic tangent, which is an infinitely differentiable function. Utilizing such an activation function is needed because otherwise the network would be unable to compute higher-order derivatives via automatic differentiation, which would prevent it from properly computing the loss in Eq. \eqref{eq: eq_loss}. As for the optimizer, we utilized ADAM (ADaptive Moment Estimation) \cite{Adam}. Lastly we want to address the metrics used to compute the losses. It is not necessary to compute all the partial losses using the same metric. Instead, we utilized Sum of Absolute Errors for the losses enforcing the physical constraints of the system: the boundary conditions loss and the normalization loss. On the other hand we utilized Mean Squared Error for the integral, differential equation, orthogonality and symmetry losses. The energy minimization loss is just computed with Eq. \eqref{eq:loss_en}. We choose to utilize SAE instead of the more standard MSE for the constraints in order to avoid the loss vanishing two quickly as the neural network outputs get closer to the desired output. This ensures that these conditions are fulfilled, making the rest of the training valid. This architecture is summarized in Figure \ref{fig:graph}}. 

\subsection{Weights} \label{sec: weights}
To allow the training to converge, we need to set the correct weights for each of the losses. Empirically, it seems that the most effective weights give each of the losses a scale, causing the network to try to reduce them in order. In particular, the chosen weights are such that the model is first forced to enforce the physical constraints of the system, such as the normalization, and only afterward it will start to try to minimize the differential equation loss. We obtain the results reported below in Table \ref{tab:weights}: 

\begin{table}[ht]
    \centering
    \begin{tabular}{c|c}
    \hline
     Loss&Weight  \\
     \hline\hline
      Normalization   & 500\\
      \hline
      Integral & 1000\\
      \hline
      Boundary conditions & 10 \\
      \hline
      Symmetry & 1000 \\
      \hline
      Orthogonality & 500\,n\\
      \hline
      Differential equation (starting) & 1\\
      \hline
      Energy minimization (starting) & 100
    \end{tabular}
    \caption{Weights for the different losses}
    \label{tab:weights}
\end{table}

This choice of weights ensures that first of all the network learns how to compute the integral. Once it has learned how to, it can use this integral to enforce normalization, which must be the first thing the network learns so that it does not just predict constant zeros and get stuck in a local minimum. After that, it will enforce the boundary and orthogonality conditions. Furthermore, the differential equation's weight is linearly increased to ensure that the solution of the actual equation takes priority in later epochs.
The symmetry loss weight has been set so high in order to guarantee that it limits the possible solutions accepted by the network and thus accelerates convergence. 

The Orthogonality loss is proportional to the quantum number of the excited state we want to study. 
\cor{See also \cite{competing_losses} for a similar approach.}

\subsection{Evaluation metrics}
When the exact solutions are known, the solution given by the trained network is evaluated by utilizing two metrics. To gauge the eigenvalue's correctness we utilize the relative error for the energy:
\begin{equation}\label{eq: rel_error_e}
    err_E = \frac{E_{Ex} - E_{PINN}}{E_{Ex}},
\end{equation}
where \(E_{Ex}\) is the analytic value of the energy, while \(E_{PINN}\) is the one predicted by the network\footnote{\cor{Here we do not use a norm and retain the sign because it provides additional information about the nature of the error. Namely, a positive $err_E$ means that the PINN underestimates the energy, while a negative $err_E$ signals an overestimation of the energy.}}.
On the other hand to evaluate the eigenvector the fidelity between the predicted wave function and the exact one \cite{fid1, fid2, fid3} is employed:
\begin{equation}\label{eq: sim_wf}
    \mathcal{F}_\psi = |\braket{\psi_{Ex}}{\psi_{PINN}}{}|^2, 
\end{equation}
 numerically evaluated as:
 \begin{equation}\label{eq: sim_wf, discrete}
     \mathcal{F}_\psi = |\sum_i \psi_{PINN}(x_i)\overline{\psi_{Ex}(x_i)}|^2
 \end{equation}
where the {\(x_i\)} are a set of points \cor{on an evenly spaced grid} spanning the whole domain, and \(\psi_{Ex}(x)\) and \(\psi_{PINN}(x)\) are the values of the wave function at the point $x$ as given by the analytic expression and by the network, respectively. Both of the vectors will be normalized before calculating the similarity, to avoid getting incorrect results due to inaccurate scaling.
This should equal one when the two wave functions are the same.
These metrics, however, might depend on the choice of evaluation points. Therefore once the training has converged we compute them $1000$ times for different sets of points, sampled in the same way we sample the training batches, to make sure that we always get a different set of points, and that each of these sets still spans the whole domain, to obtain a valid estimate of the fidelity. The final values will be the average of these estimates and will come with a standard deviation quantifying the dependence on the choice of evaluation points. For a deeper analysis, in which we discuss better the variance of these results we refer the reader to Section \ref{sec:ener}.

\subsection{Choice of the domain}
We need to choose an $L$ such that the wave function is vanishingly small at the boundaries, but at the same time, $L$ should be as small as possible to avoid training the network on areas of just zeros, which would make the network vastly underestimate the orthogonality loss \eqref{eq: orth_loss} when using Mean Squared Error. This means that the optimal choice of $L$ depends on the state that we want to train the network to predict, since for this type of potential higher energy states will tend to be less localized and we must therefore choose a domain that grows as $n$ increases\footnote{If this means that we need to extrapolate, for instance when obtaining the values of \(\psi_i(x)\) for the orthogonality loss, we just predict 0 outside the original computational domain since the wave function should be negligible outside it.}. In particular, we empirically find that a good value for $L$ for the unperturbed oscillator is:
\begin{equation} \label{eq: L}
    L(n) = 7 + 2n
\end{equation}
However, the potential of the anharmonic oscillator will grow faster than the purely harmonic one. Therefore, the resulting wave functions will be more localized, meaning that it will be better to train the network in a smaller domain. In particular, a good starting point is the interval \([-\frac{L_a}{2}, \frac{L_a}{2}]\) such that \(V_{ANH}(\frac{L_a}{2}) = V_{H}(\frac{L}{2})\), where \(V_{ANH}\) is the anharmonic potential \(\frac{1}{2} \omega^2 x^2 +\lambda x^4\) and \(V_{H}\) is the harmonic potential \(\frac{1}{2} \omega^2 x^2\). Thus, given $n$ and \(\lambda\), \(L_a\) will be the solution to:
\begin{equation} \label{eq: to_find_L}
    \lambda (\frac{L_a(n, \lambda)}{2})^4 + \omega (\frac{L_a(n, \lambda)}{2})^2 = \omega (\frac{L(n)}{2})^2 
\end{equation}

To that we add a constant \(\epsilon = 1\) on each side of the interval to ensure some leeway\footnote{But if this makes \(L_a(n, \lambda) > L(n)\), we just use \(L(n)\).}, thus obtaining:
\begin{equation} \label{eq: L, anharmonic}
    L_a = 2\left(\sqrt{\frac{-1 + \sqrt{1 + 4\frac{\lambda}{\omega}(\frac{L}{2})^2}}{2\frac{\lambda}{\omega}}} + \epsilon\right)
\end{equation}

\subsection{Materials}
\cor{All computations have been performed on a MSI GF65 Thin 10UE laptop with a NVIDIA GeForce RTX 3060 Laptop GPU}

\section{Results for the harmonic oscillator} \label{sec: harm_osc}
We train the network to predict the first six stationary states. The training is stopped as soon as the total loss becomes less than \cor{$5\times10^{-2}$} and the differential equation loss is smaller than \cor{$4\times10^{-4}$}, in order to obtain good guesses for the energies and eigenstates without requiring an overly long computational time. We set \(E_{init} = 0\) and \(a = 0.8\) in \eqref{eq:loss_en}, and obtain the values reported below in \cor{Table \ref{tab:results_unpert}}:
\begin{table}[ht]
\label{Tab_erHO}
    \centering
    \begin{tabular}{c|c|c}
    \hline
     n&\(err_E\)&\(\mathcal{F}_\psi\)  \\
     \hline\hline
        0 & $3.76\times10^{-4}$ & $0.9999665$\\
      \hline
        1 & $5.59\times10^{-4}$ & $0.9999776$\\
      \hline
      2 & $-4.44\times10^{-5}$ & $0.9998280$\\
      \hline
      3 & $1.18\times10^{-5}$ & $0.9999918$\\
      \hline
      4 & $1.99\times10^{-5}$ & $0.9999005$\\
      \hline
      5 & $5.10\times10^{-5}$ & $0.9999357$\\
    \end{tabular}
    \caption{Results for the error and the fidelity between the exact state and the state computed by the neural network for the harmonic unperturbed oscillator.}
    \label{tab:results_unpert}
\end{table}


\begin{figure*}[htbp]
    \centering
    \begin{subfigure}[b]{0.31\columnwidth}
        \includegraphics[width=\linewidth]{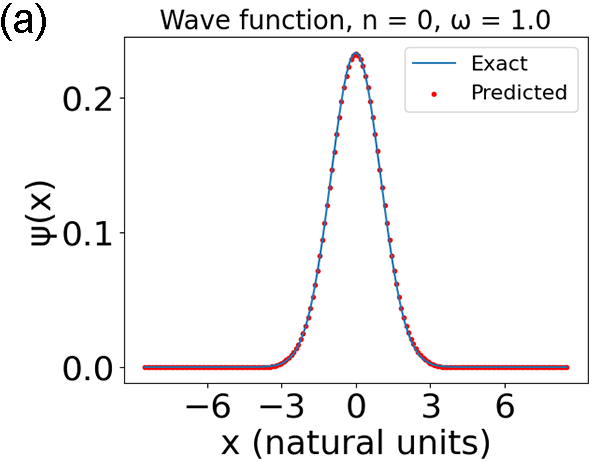}
    \end{subfigure}
    \begin{subfigure}[b]{0.31\columnwidth}
        \includegraphics[width=\linewidth]{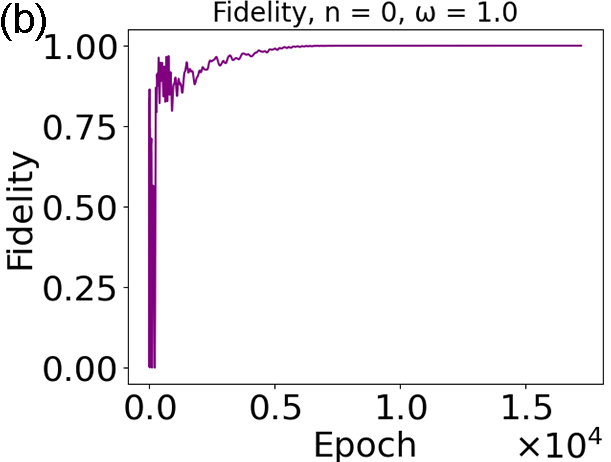}
    \end{subfigure}
    \begin{subfigure}[b]{0.31\columnwidth}
        \includegraphics[width=\linewidth]{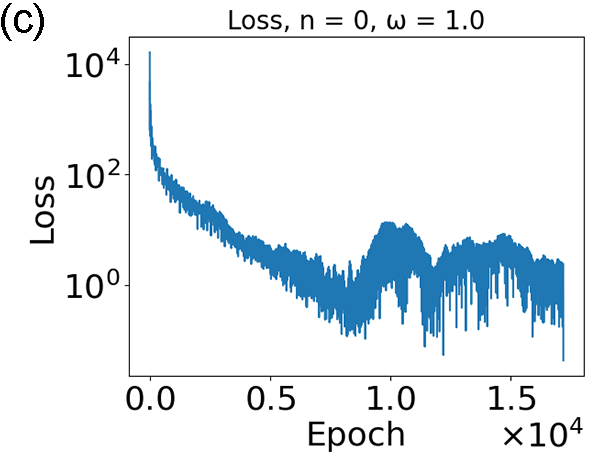}
    \end{subfigure}
    \caption{\cor{(a) Ground state for the harmonic oscillator. The red dots are the PINN's predictions, while the blue line is the \textit{ground truth}. (b) Fidelity throughout training for the ground state. (c) Loss behavior throughout training. The y axis is in logarithmic scale, therefore the oscillations on the y axis for low losses are overemphasized} }
    \label{fig:wf harm gs}
\end{figure*}






\begin{figure*}[htbp]
    \centering
    \begin{subfigure}[b]{0.31\columnwidth}
        \includegraphics[width=\linewidth]{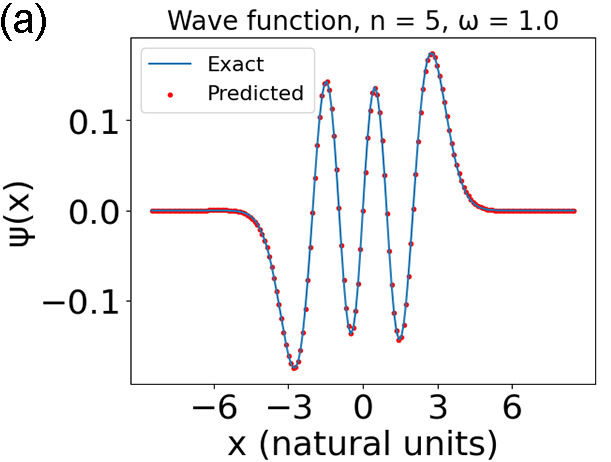}
    \end{subfigure}
    \begin{subfigure}[b]{0.31\columnwidth}
        \includegraphics[width=\linewidth]{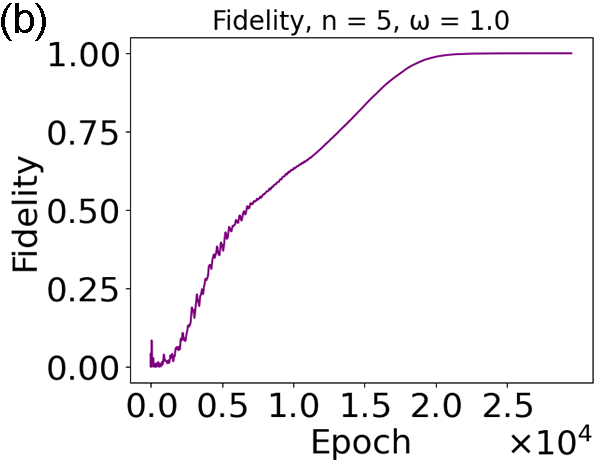}
    \end{subfigure}
    \begin{subfigure}[b]{0.31\columnwidth}
        \includegraphics[width=\linewidth]{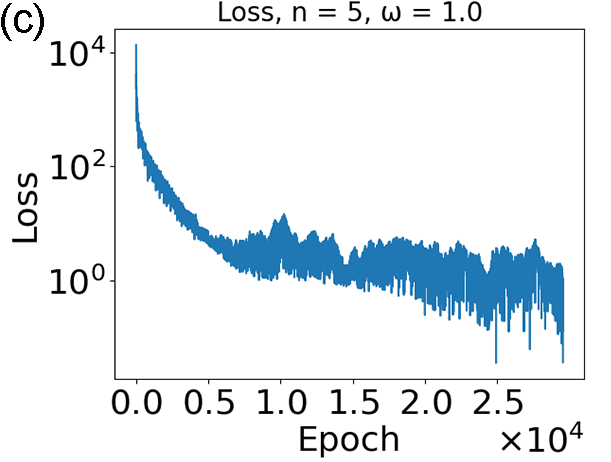}
    \end{subfigure}
    \caption{(a) Fifth excited state for the harmonic oscillator. The red dots are the PINN's predictions, while the blue line is the \textit{ground truth}. (b) Fidelity throughout training for the ground state. (c) Loss behavior throughout training. The y axis is in logarithmic scale, therefore the oscillations on the y axis for low losses are overemphasized }
    \label{fig:wf harm n5}
\end{figure*}

\begin{figure*}[htbp]
    \centering
    \begin{subfigure}[b]{0.31\columnwidth}
        \includegraphics[width=\linewidth]{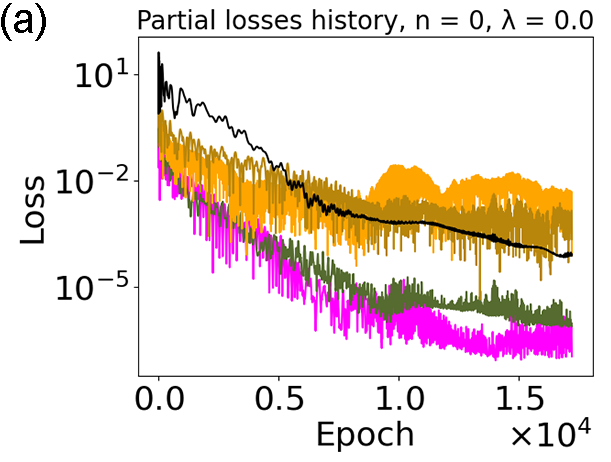}
    \end{subfigure}
    \begin{subfigure}[b]{0.31\columnwidth}
        \includegraphics[width=\linewidth]{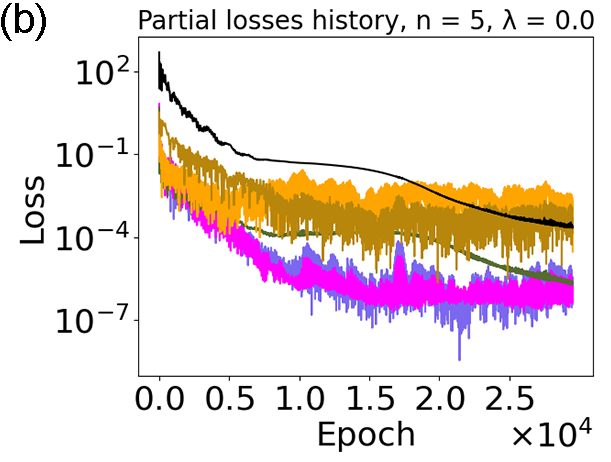}
    \end{subfigure}
    \begin{subfigure}[b]{0.31\columnwidth}
        \includegraphics[width=\linewidth]{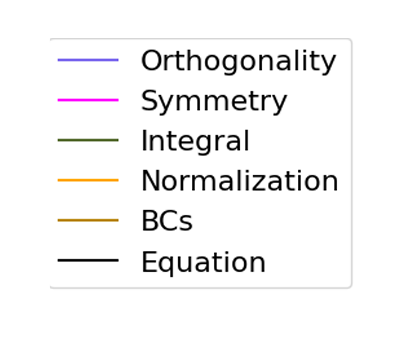}
    \end{subfigure}
    \caption{\cor{Behavior of the seven losses throughout training for the (a) ground state and (b) fifth excited state of the harmonic oscillator. The orthogonality loss is displayed in blue. Note that this loss cannot be seen in (a) as it is always $0$ for the ground state. The symmetry loss is displayed in magenta. The integral loss is displayed in olive green. The normalization loss is displayed in orange. The boundary conditions loss is displayed in brown. The differential equation loss is displayed in black}}
    \label{fig:part losses}
\end{figure*}






The error on the eigenenergy always stays beneath the 1\%. The results for the fidelity are also more than acceptable, with a distance of at most in the order of \(10^{-4}\), even though the fidelity slightly decreases as the energy increases. Some examples of the final results and the behavior of the loss and fidelity\footnote{\cor{After each training step, the \textit{ground truth} and the current PINN output are computed on a grid, and the fidelity between the two is computed. Note that this is a validation step that happens outside the training step, due to the unsupervised nature of the method. The network never actually utilizes the \textit{ground truth}}.} throughout the training process are reported in Figure \ref{fig:wf harm gs}  for the ground state, and in Figure \ref{fig:wf harm n5} for the fifth excited state. In particular, looking at the behavior of the loss \cor{(Figures \ref{fig:wf harm gs}c and \ref{fig:wf harm n5}c)} we can see it is mostly monotonically decreasing, but for $n = 5$ it takes roughly double the epochs then $n = 0$ to get to the same value. Still, it shows good convergence. Concerning fidelity \cor{(Figures \ref{fig:wf harm gs}b and \ref{fig:wf harm n5}b)}, it displays excellent behavior for $n = 5$, asymptotically approaching 1. On the other hand, the behavior for $n = 0$ is more complex. It rapidly approaches $0.8$, then goes back down to almost zero, and eventually rises up until it reaches 1. A possible explanation could be that the network was approaching the ground state, then it risked converging to an excited state, which will be orthogonal to the ground and therefore with fidelity zero, but after several epochs it managed to converge to the correct state, thanks to the energy minimization condition. \cor{Furthermore, we can qualitatively see from Figures \ref{fig:wf harm gs}a and \ref{fig:wf harm n5}a that the neural network predictions seem to line up with the \textit{ground truth}, as expected from the good values for the fidelity in Table \ref{tab:results_unpert}. It is also interesting to look at the behavior of the six\footnote{We exclude the energy minimization loss from this analysis, since it is only relevant in the initial epochs to avoid having the network converge to the wrong state.} partial losses, in Figures \ref{fig:part losses}. The overall behavior of the losses follows with good agreement the one of the total loss in Figures \ref{fig:wf harm gs}c and \ref{fig:wf harm n5}c. However, looking at the partial losses gives some important information. First of all, the differential equation and integral losses seem to oscillate far less compared to the other losses. This is something that must be taken into account when setting up the neural network's parameters, especially the learning rate. Secondly, the losses relating to the physical constraints of the system seem to start stagnating after some epochs. This is not unexpected, since the heuristic detailed in Sec. \ref{sec: weights} sets up the weights in such a way that those conditions are satisfied before the network focuses on optimizing the differential equation.}

\begin{figure*}[htbp]
    \centering
    \begin{subfigure}[b]{0.31\columnwidth}
        \includegraphics[width=\linewidth]{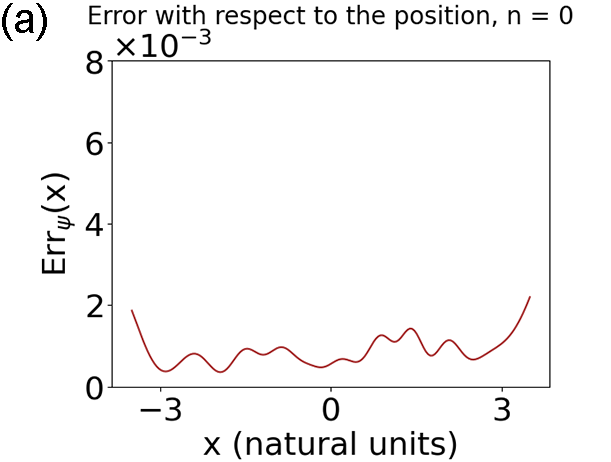}
    \end{subfigure}
    \begin{subfigure}[b]{0.31\columnwidth}
        \includegraphics[width=\linewidth]{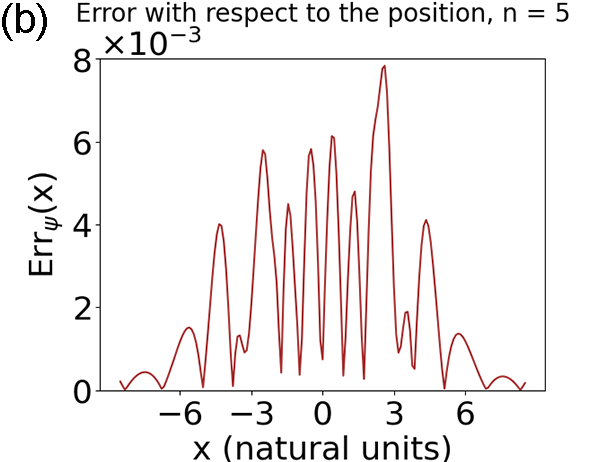}
    \end{subfigure}
    \caption{\cor{Squared error relative to the position $x$ for the (a) ground state and (c) fifth excited state.}}
    \label{fig:errors_harm}
\end{figure*}

\cor{Lastly, we can also glean useful information by looking at the absolute error with respect to the position:

\begin{equation}
    Err_{\psi}(x) = |\psi_{Ex}(x) - \psi_{PINN}(x)|
\end{equation}
This error is displayed in Figure \ref{fig:errors_harm}. In particular, for the ground state in Figure \ref{fig:errors_harm}a we can see that the error tends to be mostly symmetric with respect to $0$. This is probably a direct consequence both of the introduction of the symmetry loss and of the symmetric shape of the wavefunction itself. The same can be said for the fifth excited state in Figure \ref{fig:errors_harm}b, with the difference that the scale of the error is much greater than that of the ground state. Time-wise, training a PINN takes $20$ to $40$ minutes for the ground state, and up to some hours for the higher excited states. This increase in computational time comes from both higher complexity for the shape of the excited states and the increase in the number of dot products that we need to compute in order to obtain the orthogonality loss.}

\section{Spectrum of the anharmonic oscillator}
In this section, we report the main results of this paper. As introduced in Section \ref{sec:back}
both the cases of real and imaginary frequency are analyzed. 

 \label{sec:anharm}

\subsection{Oscillator with real frequency}
We now add the quartic potential to the oscillator. In particular, we will use \(\omega = 1\) and train the neural network with \(\lambda = 0.005 \times 2^n\), with $n$ from zero to $12$.
These choices of \(\lambda\) will allow us to study cases in which the quartic potential is just a perturbation, in which the two potentials have the same scales and in which the quartic part is dominant, up until the point where the system is basically just a pure quartic oscillator. 
In order to guarantee convergence, we exploit the results of section (\ref{sec: harm_osc}) and employ transfer learning: we initialize the weight of the network to the ones that give the eigenfunction and eigenvalue of the harmonic oscillator, thus ensuring that the network starts close to the minima. Then, as \(\lambda\) increases, we always initialize the network to the weight and biases as the already solved one with the closest \(\lambda\). Due to this initialization, it becomes less important to enforce the orthogonality and energy minimization losses, since the minima closer to the starting state should already be the one corresponding to the correct eigenstate and eigenenergy, which means that we do not need these two losses as much to drive the network away from unwanted minima. Therefore, their starting weights have been decreased to \cor{$30n$ and $0$} respectively. \cor{Furthermore, to ensure accurate results for this more complex system the convergence condition for the differential equation loss has been mede stricter, going from $4\times10^{-4}$ to $2\times10^{-5}.$} The network manages to reach the convergence conditions for all values of \(\lambda\) and $n$. Thanks to transfer learning, these results were also reached much quicker. \cor{Time wise, it takes $5$ to $10$ minutes for lower excited states, compared to $20$ to $40$ minutes when given a clean initialization. Likewise, the convergence condition is reached much faster, with the harmonic oscillator requiring around $15000$ epochs starting from a clean initialization, while tha anharmonic oscillator with $\lambda = 20.48$ takes only around $2000$ epochs.}

\begin{table}
    \centering
    \begin{tabular}{c|c|c}
    \hline
     n&\(\lambda\)&\((E_{pert} - E_{PINN}) / E_{pert} \)  \\
     \hline\hline
        0 & 0.005 & $2.06\times10^{-2}$\\
      \hline
        5 & 0.005 & $-8.67\times10^{-4}$\\
    \hline
    2 & 0.16 &$ 1.56\times10^{-2}$\\
      \hline
      2 & 0.32 & $-1.87\times10^{-3}$ \\
      \hline
      2 & 0.64 &$ -4.79\times10^{-2}$ \\
      \hline
      3 & 1.28 & $-2.41\times10^{-1}$ \\
      \hline
      0 & 10.24 &$ -4.18\times10^{-1}$ \\
      \hline
      5 & 10.24 & $-7.94\times10^{-1}$ \\
    \end{tabular}
    \caption{Some results for the anharmonic oscillator. We can clearly see how for low \(\lambda\) the energy predicted in perturbation theory and the one predicted by the neural network agree, but differ substantially from \(\lambda\)s comparable to or higher than \(\omega\)}
    \label{tab: en_discrepancies}
\end{table}

\begin{figure*}[htbp]
    \centering
    \begin{subfigure}[b]{0.31\columnwidth}
        \includegraphics[width=\linewidth]{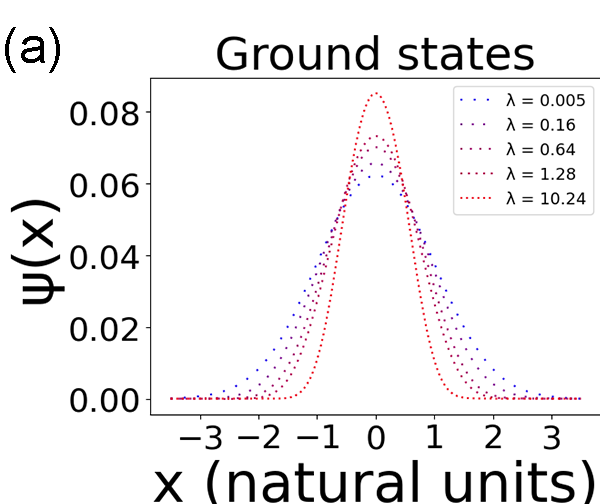}
    \end{subfigure}
    \begin{subfigure}[b]{0.31\columnwidth}
        \includegraphics[width=\linewidth]{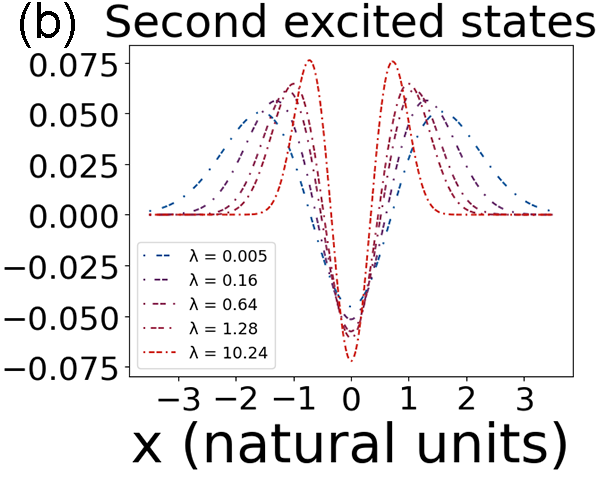}
    \end{subfigure}
    
    \begin{subfigure}[b]{0.31\columnwidth}
        \includegraphics[width=\linewidth]{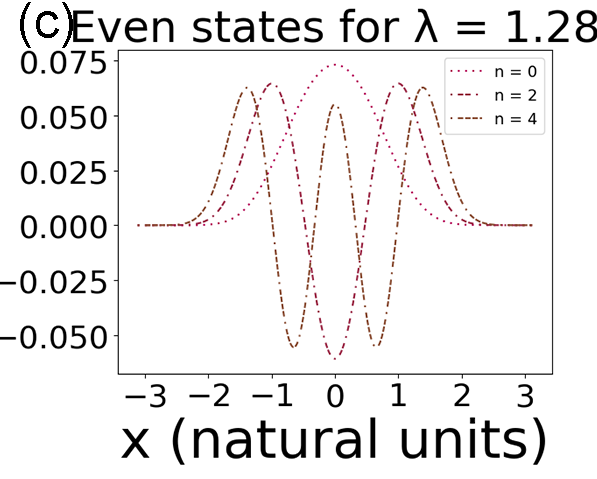}
    \end{subfigure}
    \begin{subfigure}[b]{0.31\columnwidth}
        \includegraphics[width=\linewidth]{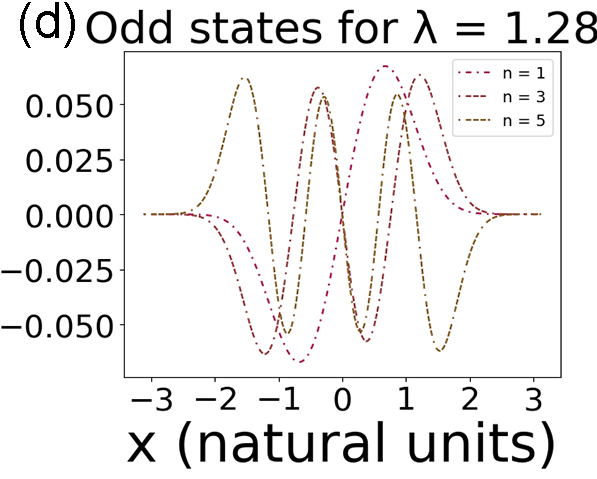}
    \end{subfigure}
    \caption{(a) Ground state and (b) second excited state of the anharmonic oscillator for \(\omega = 1\); \(\lambda = 0.005,\,0.16,\,0.64,\,1.28,\) and \(10.24\). The color of the wave functions goes from blue to red as \(\lambda\) increases. (c) Even and (d) Odd stationary states of the anharmonic oscillator for \(\omega = 1\); \(\lambda = 1.28\). The number of dashes between each dot in the plot corresponds to the value of $n$ of that eigenstate.}
    \label{fig: plots_anharm}
\end{figure*}

 At this point we compare the obtained eigenenergies with those one gets from the standard perturbation theory. In particular, we will have, for \(\omega = 1\) \cite{hioe1976}:
\begin{dmath} \label{Osc pert}
    E_{n, \lambda} = (n + \frac{1}{2}) + \frac{3}{4}\lambda( 1 + 2n(n + 1)) - \lambda^2 \left(\frac{(n + 1) (n + \frac{3}{2}) ^ 2  (n + 2)}{2 + 3 \lambda (2  n + 3)} - \frac{n (n - \frac{1}{2}) ^ 2  (n - 1)}{2 + 3 \lambda  (2  n - 1)} +  \frac{(n + 1) (n + 2)  (n + 3)  (n + 4)}{16 (4 + 6 \lambda  (2  n + 5))} - \frac{n (n - 1)  (n - 2)  (n - 3)}{16  (4 + 6  \lambda (2  n - 3))}\right)
\end{dmath}
As can be seen in table \ref{tab: en_discrepancies}, the results agree for low \(\lambda\) but diverge as the contribution of the quartic factor increases, which is exactly the behavior we expected. \cor{Note that in this case the energy computed in perturbation theory does not constitute a \textit{ground truth}, and the discrepancy of up to $79\%$ does not indicate poor performance of the network. This discrepancy comes from the fact that the assumptions of perturbation theory do not hold for such large values of $\lambda$. This leads to $E_{pert}$ to be largely incorrect while $E_{PINN}$ is probably much closer to the real value. To check the correctness of the results, we compare them to those obtained by the Legendre mesh method \cite{Leg_mesh_1}, which is known to display uncommonly high accuracy for the eigenergies of the anharmonic oscillator \cite{leg_mesh_2}. In particular, we utilize the implementation in the LagrangeMesh Mathematica package \cite{del_Valle_2023}. Some of the results are summarized in Table \ref{tab: en_anharm}. The discrepancy remains below the $1\%$, confirming the efficacy of the model. It is important to note that utilizing Lagrange Mesh is both faster and more accurate then training a PINN. Utilization of PINNs should, however, show advantages for more complex systems, due to the expressive powers of neural networks. This analysis aims to be a building block towards these more complex applications. Another future application could be to train a network that takes as input both $x$ and $\lambda$, and outputs the $\psi(x)$ and $E$ for the anharmonic oscillator with that value of $\lambda$. }
\\
A comparative plot of the obtained data is reported in Figure \ref{fig: plots_anharm}. First, we take a look at the dependence on \(\lambda\) of the ground state and the state $n = 2$ (Figures \ref{fig: plots_anharm}a and \ref{fig: plots_anharm}b). The general shape of the states remains the same, however as \(\lambda\) increases the state becomes more localized since the potential will grow more rapidly. Likewise, it is interesting to look at all the states for a given \(\lambda\) (Figures \ref{fig: plots_anharm}c, \ref{fig: plots_anharm}d). Once again they have the same general shape as the harmonic states, as expected.

\begin{table}
    \centering
    \begin{tabular}{c|c|c|c}
    \hline
     n&\(\lambda\)&\(err_E \)&\(\mathcal{F}_\psi\)  \\
     \hline\hline
        0 & 0.005 & $-5.78\times10^{-4}$ & $0.9999970$\\ 
      \hline
        5 & 0.005 & $5.78\times10^{-6}$ & $0.9999962$\\ 
    \hline
    2 & 0.16 &$2.29\times10^{-3}$ & $0.9999969$\\ 
      \hline
      2 & 0.32 & $-6.28\times10^{-4}$ & $0.9999982$ \\ 
      \hline
      2 & 0.64 &$-8.25\times10^{-5}$ & $0.9999972$ \\ 
      \hline
      3 & 1.28 & $3.10\times10^{-4}$ & $0.99999921$\\ 
      \hline
      0 & 10.24 &$-2.74\times10^{-4}$ & $0.999999974$ \\ 
      \hline
      5 & 10.24 & $-1.80\times10^{-6}$ & $0.9999999983$ \\ 
    \end{tabular}
    \caption{\cor{Results for the error on the energy between the state computed by the Lagrange Mesh method and the one computed neural network for some values of $n$ and $\lambda$ of the anharmonic oscillator.}}
    \label{tab: en_anharm}
\end{table}

\subsection{Oscillator with imaginary frequency}

Lastly, setting an imaginary value for \(\omega\) or, equivalently, taking \(r = w^2 < 0\), will allow us to study a potential that is used as a model for tunneling and bond formation in diatomic molecules. \\


Qualitatively, we can see from the shape of the potential that we can expect the ground state to be a bound state. Therefore, we set a negative value for \(E_{init}\) in the condition given in Eq.\eqref{eq:loss_en} to make it easier for the network to find negative energy states.
For the potential parameters, we start with \(\lambda = 1\), \(w^2 = -14\). Once the network has been trained for these parameters, we will be able to more easily find the solution for any other set of parameters by using transfer learning, as we did in the previous section.

\begin{figure}
    \centering
    \includegraphics[width= 0.49 \linewidth]{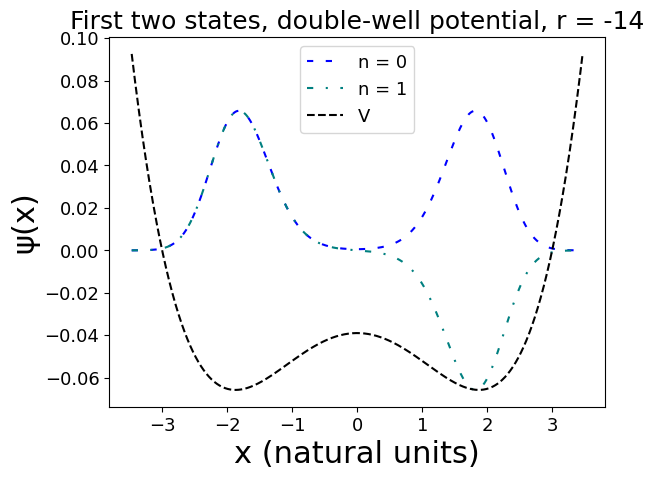}
    \caption{Wave function for the ground state (blue sparse dashed line) and first excited state (blue-green dashedotted line) of the double-well. The potential (black dense dashed line) has been centered and scaled for better visualization, and should only be taken qualitatively}
    \label{fig:gs symm}
\end{figure}

\begin{figure*}[htbp]
    \centering
    \begin{subfigure}[b]{0.31\columnwidth}
        \includegraphics[width=\linewidth]{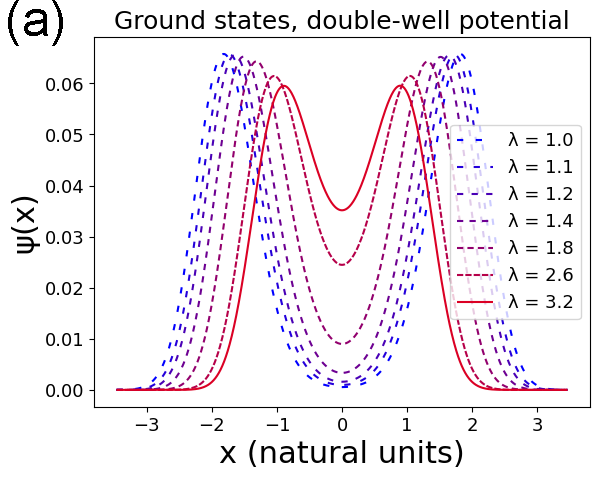}
    \end{subfigure}
    \begin{subfigure}[b]{0.31\columnwidth}
        \includegraphics[width=\linewidth]{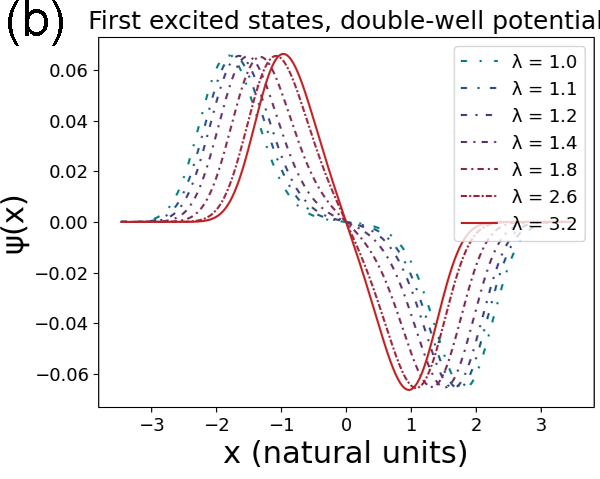}
    \end{subfigure}
    \caption{(a) Ground state and (b) second excited state of the double-well for
\(r = -14\); \(\lambda =\{ 1, 1.1, 1.2, 1.4, 1.8, 2.6, 3.2 \}\). The color
of the wave functions goes from blue to red as \(\lambda\) increases. Furthermore, the spacing between each dash becomes tighter the higher \(\lambda\).}
    \label{fig: plots_double_well}
\end{figure*}

For these given parameters, the energies of the two states are extremely close. As such, we had to impose a stricter convergence condition \cor{compared to the harmonic oscillator}, stopping the training when the differential equation loss was less than $2\times10^{-5}$, \cor{as we did for the anharmonic oscillator with real frequency}. The network converged for all the \cor{addressed} states. It took \cor{after around $64000$ epochs for the ground state and around $112000$ epochs for the first excited state}. We obtained an energy estimate of $-9.6808$ for the ground state and $-9.6806$ for the first excited state, \cor{the same values that can be obtained by the Lagrange Mesh method.} As expected, these energies are negative as they correspond to bound states. Furthermore, by increasing \(\lambda\) the eigenergies get closer to 0, as the potential becomes closer to a pure quartic oscillator and, therefore, the states get closer to unbound states. It is also useful to look at the shape of these eigenstates (Figure \ref{fig:gs symm}). These are, once again, what we expect from the shape of the potential: two peaks corresponding to the minima in the potential. The ground and the first excited state are also extremely similar, except for one symmetric and one anti-symmetric, and the peak with the same sign almost perfectly overlaps, thus justifying the closeness of the energy levels. \cor{The correctness of these eigenfunctions can also be confirmed by looking at their fidelity with the ones computed by the Lagrange Mesh method. The values of the fidelity are $0.99999984$ and $0.99999986$ for the ground state and first excited state, respectively.} \cor{We also once again utilized transfer learning to train the network with different values of $\lambda$, and} it is interesting to note, from Figure \ref{fig: plots_double_well}a that the local minima of the wavefunction corresponding to $x = 0$ assumes a higher value as \(\lambda\) increases. On the other hand, in the first excited we always have a node in $x = 0$, see Figure \ref{fig: plots_double_well}b.

\section{Pure quartic oscillator} \label{sec:pureq}
\begin{figure} 
    \centering
    \includegraphics[width=0.49\linewidth]{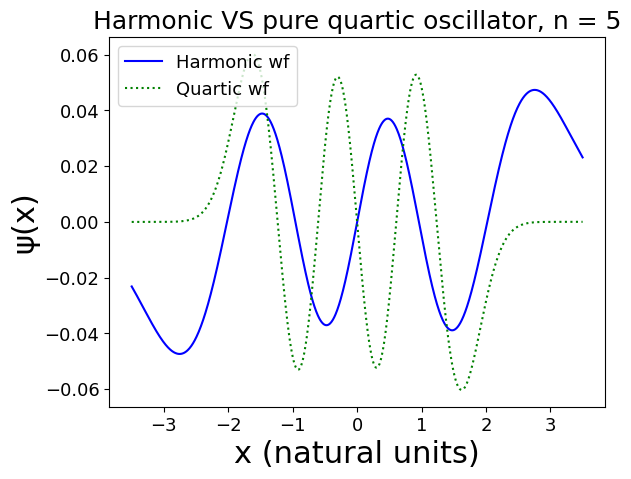}
    \caption{Comparison between the fifth excited state of the harmonic oscillator (blue solid line) and the pure quartic oscillator (green dotted line).}
    \label{fig:fifth_quartic}
\end{figure}

We now relay the results for the pure quartic oscillator, i.e.
\begin{equation}\label{eq: pure quartic}
    V(x) = \lambda x^4
\end{equation}
with \cor{\(\lambda = 1.28\)}. In order to initialize the network to a good state, we start with the weights and biases of the anharmonic oscillator with \(\omega^2 = 1\) and \cor{\(\lambda = 1.28\)}. The network converged effectively for all states \cor{given the same convergence condition as the anharmonic oscillator}. In Figure \ref{fig:fifth_quartic} we show one of the excited states of the pure quartic oscillator compared with the state at the same quantum number for the harmonic oscillator. Once again the two have the same general shape, but the quartic one is much more localized since the potential grows faster with the position $x$. \cor{We also once again check the correctness of the PINN's predictions by comparing them to those obtained with the Lagrange mesh method, and verify their excellent agreement. These results are summarized in Table \ref{tab: en_quartic}.}

\begin{table}
    \centering
    \begin{tabular}{c|c|c}
    \hline
     n&\(err_E \)& \(\mathcal{F}_\psi\) \\
     \hline\hline
        0 & $3.57\times10^{-4}$ & 0.99999984\\ 
      \hline
        1 &  $5.39\times10^{-4}$ & 0.99999977\\ 
    \hline
    2 & $3.03\times10^{-6}$ & 0.99999940\\ 
      \hline
      3 & $-9.51\times10^{-6}$ & 0.99999986 \\ 
      \hline
      4 & $-1.06\times10^{-5}$ & 0.99999955 \\ 
      \hline
      5 & $-2.19\times10^{-6}$ & 0.999999986\\ 
    \end{tabular}
    \caption{\cor{Results for the error on the energy and fidelity between the state computed by the Lagrange Mesh method and the one computed neural network for the pure quartic oscillator with $\lambda = 1.28$.}}
    \label{tab: en_quartic}
\end{table}

\section{Energy behavior}\label{sec:ener}

\begin{figure}[ht] 
    \centering
    \includegraphics[width=0.7\linewidth]{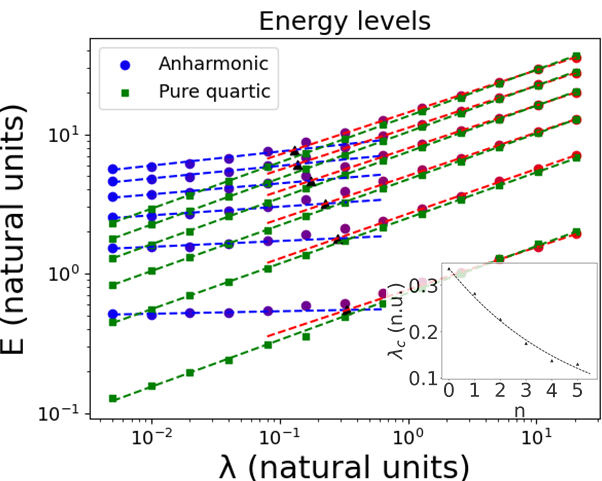}
    \caption{Energy levels of the anharmonic oscillator (blue or red circles) in comparison with the pure quartic oscillator (green squares), with exponential fits for the anharmonic eigenvalues at low \(\lambda\) (blue dashed line), anharmonic eigenvalues at high \(\lambda\) (red dashed line), pure eigenvalues at high \(\lambda\) (green dashed line). Both axes are in logarithmic scale. In the inset: A plot showing the values of \(\lambda\) corresponding to the intersections of the two exponentials corresponding to the low-\(\lambda\) and high-\(\lambda\) regimes for each energy level (\(\lambda_{c}\)). On the horizontal axis is the quantum number $n$, whereas and on the vertical axis is \(\lambda_{c}\). These values have also been fitted and show a roughly exponential behavior. The intersection points have also been signaled as black triangles in the main plot.}
    \label{fig: eigens_fit}
\end{figure}

The last result of our paper concerns the scaling behavior and the spacing of the energy levels as a function of $\lambda$. 
It is indeed interesting to check whether our method is able to assess the departure from the validity of the perturbative regime. 
We compare in Figure \ref{fig: eigens_fit} the eigenvalues of the anharmonic oscillator against the ones of the pure quartic oscillator. 
It is easy to see how the eigenvalues of the pure quartic oscillator increase at a roughly exponential rate when plotted against the logarithm of \(\lambda\) (or, equivalently, they are proportional in log-log scale). On the other hand, we can recognize two regimes for the values of the anharmonic oscillator. Both seem to be roughly exponential, but for low values of \(\lambda\) the energy increases much slower than for higher ones. Furthermore, for high \(\lambda\) the anharmonic and pure quartic eigenvalues line up, since the harmonic contribution becomes negligible. We can get a rough idea of this behavior by performing a linear fit for the logarithms of \(\lambda\) and E. In particular, we perform three fits: one for the low-\(\lambda\) region (roughly \(\lambda < 0.1\)) of the anharmonic oscillator, one for the high-\(\lambda\) region (roughly \(\lambda > 2.0\)) of the anharmonic oscillator, and lastly a fit for all the values of the quartic oscillator. We obtain, as expected, exponential behavior for each of the studied regions. Furthermore, we can see how the high-\(\lambda\) behavior of the anharmonic oscillator matches the quartic oscillator for the last values of \(\lambda\) when the quadratic part becomes negligible. See Appendix \ref{app:tables} for the numerical values of the fits. 

\section{Conclusions}
\label{sec:concl}
In this work, we utilized a novel unsupervised deep-learning-based model to find the eigenvalues and eigenstates for a potential with a plethora of applications in physics. To benchmark the performances of our method, we first started by finding the first six stationary states of the harmonic oscillator. The obtained results then served as initialization to solve the anharmonic oscillator and its pure quartic version for growing values of the parameter \(\lambda\), \cor{where it has been validated with results obtained via the Lagrangian-mesh method}. Our method is reliable on the full range of $\lambda$ and it allowed us to establish two exponential scalings for the energy levels. The result seems to suggest there is a critical value $\lambda_c$ that decreases with the energy for which the perturbative expansion is no longer reliable. This is consistent with the previous results, in fact, the spectrum does not have a constant spacing due to its nonlinear dependence
on the occupation number \cite{PRD1}. Therefore, this opens an interesting avenue of research to study if such conclusions can be drawn also for field theories in 
higher dimensions and in particular if this could be related to the phase transition that could be studied via mapping of the scalar $\lambda \phi^4$\ in $(3+1)-$dimensions to an Ising model in $d\geq 4$ dimensions. For the machine learning side, since the changes in the wave function as \(\lambda\) increases seem to be smooth it is, in principle, possible to use these results to train a neural network that can take both \(\lambda\) and $x$ as the input, and outputs the wave function at $x$ and energy for a given eigenstate of the potential with the given \(\lambda\). \cor{As a final remark, we like to stress that already established methods could be more efficient in terms of resources, our work paves the way to the applications of PINNs to complex scenario of interest in quantum science and technology, where classical numerical techniques are struggling.}

\subsection*{Code Availability}
The code producing the results in the paper is available upon reasonable request from the Authors.

\subsection*{Acknowledgements}
The Authors acknowledge the project CQES of the Italian Space Agency (ASI) for having partially supported this research (grant N. 2023-46-HH.0).

\subsection*{Author contribution statement}
L.B. and A.M. contributed equally to this work.

\newpage

\appendix
\section{Coefficients of the linear fits} \label{app:tables}
In this Section, we report the numerical coefficients of the fits in Figure \ref{fig: eigens_fit}. 
In Table \ref{tab: a_b_low} we report the interpolation coefficients used to draw the blue dashed lines in Figure \ref{fig: eigens_fit}, equivalently in Tables \ref{tab: a_b_high} and \ref{tab: a_b_quartic} we report the interpolation coefficients used to draw the red and green dashed lines in Figure \ref{fig: eigens_fit}. 
That is, the coefficients of the line:
\begin{equation}
    \log(E_n) = a \log(\lambda) + b
\end{equation}
We can see, in particular, that the values of $a$ for the low-\(\lambda\) region (\(\lambda < 0.1\)) are much lower than for the other two. On the other hand, the high-\(\lambda\) (\(\lambda > 2\)) and quartic coefficients are comparable. We expect that if we were to redo the fit with higher values of \(\lambda\) and increase the cutoff for the high-\(\lambda\) region each of the pairs at the same $n$ will become increasingly similar. At any rate, we can say that in each of the three cases \(\log(\lambda)\) and \(\log(E)\) are directly proportional.

\begin{table*}
    \begin{minipage}[t]{0.28\linewidth}
    \centering
    \begin{tabular}{c|c|c}
    \hline
     n & a & b  \\
     \hline\hline
        0 & 0.0180 & -0.580\\
      \hline
        1 & 0.0419 & 0.635\\
    \hline
    2 & 0.0621 & 1.25\\
      \hline
      3 & 0.0787 & 1.67\\
      \hline
      4 & 0.0903 & 1.99\\
      \hline
      5 & 0.101 & 2.25\\
    \end{tabular}
    \caption{The fit coefficients for the \(\lambda < 0.1\) region, anharmonic oscillator}
    \label{tab: a_b_low}
    \end{minipage}
    \quad
    \begin{minipage}[t]{0.28\linewidth}
    \centering
    \begin{tabular}{c|c|c}
    \hline
     n & a & b  \\
     \hline\hline
        0 & 0.305 & -0.266\\
      \hline
        1 & 0.321 & 0.989\\
    \hline
    2 & 0.310 & 1.62\\
      \hline
      3 & 0.307 & 2.07\\
      \hline
      4 & 0.302 & 2.41\\
      \hline
      5 & 0.303 & 2.66\\
    \end{tabular}
    \caption{The fit coefficients for the \(\lambda > 2\) region, anharmonic oscillator}
    \label{tab: a_b_high}
    \end{minipage}
    \quad
    \begin{minipage}[t]{0.28\linewidth}
    \centering
    \begin{tabular}{c|c|c}
    \hline
     n & a & b  \\
     \hline\hline
        0 & 0.336 & -0.321\\
      \hline
        1 & 0.327 & 0.919\\
    \hline
    2 & 0.329 & 1.56\\
      \hline
      3 & 0.332 & 2.01\\
      \hline
      4 & 0.332 & 2.34\\
      \hline
      5 & 0.332 & 2.60\\
    \end{tabular}
    \caption{The fit coefficients for the interpolation lines plot of the pure quartic oscillator}
    \label{tab: a_b_quartic}
    \end{minipage}
\end{table*}

In Table \ref{tab: inters_coeff} we show the values \(\lambda_c\) of \(\lambda\) for which the low-\(\lambda\) and high-\(\lambda\) fits intersect and the corresponding energies \(E_c\), which can be used as an indication for the phase transition from the low-\(\lambda\) perturbative regime to the high-\(\lambda\) quartic regime.   The critical value \(\log(\lambda_c)\) seems to be inversely proportional with respect to the $n-$th energy level, 
\(\lambda_c\) decreases as the energy of the state increases, causing an earlier deviation from the perturbative regime.

\begin{table}
    \centering
    \begin{tabular}{c|c|c}
    \hline
     n & \(\lambda_c\) & \(E_c\)  \\
     \hline\hline
        0 & 0.333 & 0.549\\
      \hline
        1 & 0.282 & 1.79\\
    \hline
    2 & 0.226 & 3.18\\
      \hline
      3 & 0.175 & 4.64\\
      \hline
      4 & 0.137 & 6.11\\
      \hline
      5 & 0.129 & 7.72\\
    \end{tabular}
    \caption{The fit coefficients for the pure quartic oscillator}
    \label{tab: inters_coeff}
\end{table}

\newpage
\bibliography{refs}
\end{document}